\documentclass[a4paper,12pt]{article}
\usepackage[pdftex]{graphicx}
\usepackage{graphicx}
\usepackage{graphics}
\usepackage{epsfig}
\usepackage{subfigure}
\usepackage{amssymb}
\usepackage{amsmath}
\usepackage{color}
\usepackage{authblk}
\usepackage{afterpage}
\usepackage[square,comma,sort&compress]{}
\usepackage{sectsty}
\sectionfont{\fontsize{12}{12}\selectfont}  
\subsectionfont{\fontsize{12}{12}\selectfont}  
\font\myfont=cmr12 at 12pt
\author[1,2]{\myfont Awaneesh Singh\thanks{awaneesh11@gmail.com}}
\author[2]{\myfont Amrita Singh}
\author[2]{\myfont Anirban Chakraborti\thanks{anirban@jnu.ac.in}}

\affil[1]{\myfont Department of Physics, Institute of Chemical Technology, Mumbai-400019, India.}
\affil[2]{\myfont School of Computational and Integrative Sciences, Jawaharlal Nehru University, New Delhi-110067, India.}
\title{\fontsize{12}{12} \bf{Effect of bond-disorder on the phase-separation kinetics of binary mixtures: a Monte Carlo simulation study}}
\date{}

\begin{document}
\maketitle

\begin{abstract}
We present Monte Carlo (MC) simulation studies of phase separation in binary ($AB$) mixtures with bond-disorder that is introduced in two different ways: (i) at randomly selected lattice sites and (ii) at regularly selected sites. The Ising model with spin exchange (Kawasaki) dynamics represents the segregation kinetics in conserved binary mixtures. We find that the dynamical scaling changes significantly by varying the number of disordered sites in the case where bond-disorder is introduced at the randomly selected sites. On the other hand, when we introduce the bond-disorder in a regular fashion, the system follows the dynamical scaling for the modest number of disordered sites. For higher number of disordered sites, the evolution morphology illustrates a lamellar pattern formation. Our MC results are consistent with the Lifshitz-Slyozov (LS) power-law growth in all the cases. 
\end{abstract}

\section{Introduction}
\label{Intro}
A binary ($AB$) mixture, which is homogeneous (or disordered) at high temperatures becomes thermodynamically unstable when rapidly quenched inside the coexistence curve. Then, the binary ($AB$) mixture undergoes phase separation (or ordering) via the formation and growth of domains enriched in either component. Much research interest has focused on this far-from-equilibrium evolution \cite{ab94,pwphi9,bf2013,aophi2,dp2004,ralj2008}. The domain morphologies are usually quantified by two important properties: a) the domain growth law (characteristic domain size $L(t)$ grows with time $t$), which depends on general system properties, e.g., the nature of conservation laws governing the domain evolution, the presence of hydrodynamic velocity fields, the presence of quenched or annealed disorder, etc. b) the correlation function or its Fourier transform, the structure factor, which is a measure of the domain morphology \cite{ab94,pwphi9}. 

There now exists a good understanding of phase separation dynamics for binary mixtures \cite{smv2011,spd2012,spd2014,skp2015,dph05,dph06}. Normally, for a pure and isotropic system, domain growth follows a power-law behavior, $L(t) \sim t^{\phi}$ where $\phi$ is referred to as the growth exponent. For the case with nonconserved order parameter (ordering of a magnet into up and down phases), the system obeys the Lifshitz-Cahn-Allen (LCA) growth law with $\phi=1/2$ \cite{ab94,pwphi9,bf2013,aophi2}. For the case with conserved order parameter (diffusion driven phase separation of an $AB$ mixture into $A$-rich and $B$-rich phases). The system obeys the Lifshitz-Slyozov (LS) growth law with $\phi=1/3$ \cite{ab94,pwphi9,bf2013,aophi2}. However, including the hydrodynamic effects in a system with conserved order parameter (e.g., segregation of a binary fluid), there appear to be various domain growth regimes, depending on the dimensionality and system parameters \cite{spd2014,kk91,pd92,adp12,kdb99}. 

In reality, the experimental systems are neither pure nor isotropic. Usually, they always endure impurities (annealed or quenched) within the system. An important set of results has been well documented from both analytical and numerical studies on phase ordering in systems with quenched disorder \cite{hh85,gs85,sg32,oc86,cgg87,pcp91,pp92,hayakawa91,bh91,ppr05,ppr04}. The quench disorder (considered as an immobile impurity) is introduced into the pure Ising model by either random spin-spin exchange interaction, i.e., random-bond Ising model (RBIM) \cite{oc86,ppr05,ppr04} or by introducing a site-dependent random-field Ising model (RFIM) \cite{nv88,bm85}. In general, sites of quenched disorder act as traps for domain boundaries with the energy barrier being dependent on the domain size. In this regard, a significant contribution is made by Huse and Henley (HH)\cite{hh85} to understand the growth law for the bond disorder case. They argued that the energy barrier follows power-law dependence on domain size: $E_b(L) \simeq \epsilon L^\psi$. Here, $\epsilon$ is the disorder strength and $\psi$ is the barrier exponent that depends on the roughening exponent $\zeta$ and the pinning exponent $\chi$ as $\psi=\chi/(2-\zeta)$; the roughening and pinning exponents are related as $\chi=2\zeta+d-3$, where $d$ is the system dimensionality. Consequently, the normal power law growth ($L(t)\sim t^\phi$) of the characteristic domain size changes over to a logarithmic growth $L(t)\sim (\ln t)^\phi$. A few numerical simulations \cite{gs85,sg32,oc86,cgg87,pcp91,pp92,hayakawa91,bh91,ghs95} and experiments \cite{iei90, lla2000, llo2001} were performed to test the HH proposal. Nevertheless, to date, no definite confirmation of logarithmic growth in the asymptotic regime is observed. 

Later, Paul, Puri, and Rieger (PPR) \cite{ppr05,ppr04} reconsidered this problem via extensive Monte Carlo (MC) simulations of the RBIM with nonconserved (Glauber) spin-flip kinetics, and conserved (Kawasaki) spin exchange kinetics. In contrast to HH scenario, PPR observed the normal power-law domain growth with temperature and disorder dependent growth exponent, similar to the one seen in the experiments \cite{iei90, lla2000, llo2001} on domain growth in the disordered system. PPR proposed that the growth exponents can be understood in the framework of a logarithmic domain size dependence of trapping barrier ($E_b(L) \simeq \epsilon \ln(1+L)$) rather than power-law \cite{ppr05}. At early times, domains coarsening is not affected by disorder due to small energy barriers, and therefore, the system evolves like a pure system. At late times, the disorder traps become effective at a crossover length scale, and it can only move by thermal activation over the corresponding energy barrier. Thus, thermal fluctuations drive the asymptotic domain growth in disordered systems \cite{ppr05,ppr04}. This should be contrasted with the pure case, where thermal fluctuations are irrelevant. In these cases, quench disorder was introduced by uniformly varying the strength of the spin-spin exchange interaction between zero and one at all the lattice sites.

In this paper, we present MC simulation of domain coarsening in binary mixtures with quenched disorder using conserved (Kawasaki) spin-exchange kinetics. Here, we introduce the disorder in two different ways: a) at randomly selected lattice sites and b) at regularly selected lattice sites. We consider the strength of the spin-spin exchange interaction equal to zero at these selected sites (equivalent to have sites at $T\gg T_c$ called disordered sites) and equal to one at the rest of the sites. By varying the number of selected sites, we discuss the effect of disorder on the domain growth law and the dynamical scaling. Our simulations are aimed to gain a conceptual understanding of these disordered systems where theoretical calculations are challenging at present. This paper is organized as follows. In Sec. \ref{Method}, we describe the methodology we used to simulate the system. In Sec. \ref{Nr}, we present the results and discussion for both the cases of introducing disorder. Finally, Sec. \ref{Conclusions} concludes this paper with the summary of our results. 

\section{Methodology}
\label{Method}
Let us start with a description of Monte Carlo (MC) simulations for the study of phase separation in binary ($AB$) mixtures. The Hamiltonian for the Ising system is described by 
\begin{equation}
 H=-\sum_{<ij>}J_{ij}S_{i}S_{j}, \qquad S_{i}=\pm 1. 
\label{ham}
\end{equation}
Here, $S_{i}$ denotes the spin variable at site $i$. We consider two state spins: $S_{i}=+1$ when a lattice site $i$ is occupied by an $A$ atom and $S_{i}=-1$ when occupied by a $B$ atom. The subscript $<ij>$ in Eq.~\ref{ham} denotes a sum over nearest-neighbor pairs only. The term $J_{ij}$ denotes the strength of the spin-spin exchange interaction between nearest-neighbor spins. We consider the case where $J_{ij}\geq 0$ so that the system is locally ferromagnetic. The case where a system has both $J_{ij}\geq 0$ (ferromagnetic) and $J_{ij}\leq 0$ (antiferromagnetic) is relevant to spin glasses. Normally, in MC simulation for a pure phase-separating binary $(AB)$ mixture, we consider $J_{ij}=1$ with a critical temperature $T_c\simeq2.269/k_B$ for a $d=2$ square lattice. Further, $J_{ij}=0$ corresponds to the maximally disordered system, equivalent to the system at $T\gg T_c$ where all proposed spin exchanges will be accepted. 

In our MC simulation, spins are placed on a square lattice $(L_x \times L_y)$ with periodic boundary conditions in both the directions. We assign random initial orientations: up ($S_{i}=+1$) or down ($S_{i}=-1$) to each spin and rapidly quench the system to $T<T_c$. The quench disorder is introduced via exchange coupling as $J_{ij}=1-\epsilon$, where $\epsilon$ quantifies the degree of disorder. In this paper, we considered only two values of the degree of disorder, $\epsilon =0$ (pure system) and $\epsilon =1$ (disordered sites corresponding to impurities in the system). Notably, in  PPR's study \cite{ppr05} $J_{ij}$ is uniformly distributed in the interval [$1-\epsilon,1$], where the limit $\epsilon=0$ corresponds to the pure case and $\epsilon=1$ corresponds to the maximally disordered case with $J_{ij}\in [0,1]$. 

We perform our MC simulations for two different cases corresponding to the way we introduce disorder into the system. In Case 1 we randomly selected a fraction of sites with $\epsilon =1$ and in Case 2, we picked the same fraction of sites in a regular fashion. The remaining lattice sites are set to $\epsilon =0$. Shortly, we present the results for three different percentages of disorder sites ($\epsilon =1$) namely at 2\%, 5\% and 10\% of total sites, $N$, for both the cases and compare them with the pure case ($\epsilon =0$). The initial condition of the system corresponds to a critical quench with 50\% $A$ (up) and 50\% $B$ (down) spins.

We place the Ising system in contact with a heat bath to associate stochastic dynamics. The resultant dynamical model is referred to as a \textit{Kinetic Ising Model}. We consider spin-exchange (Kawasaki) kinetics, an appropriate model to study the phase separation in $AB$ mixtures \cite{pwphi9, smv2011}. It is straight forward to implement MC simulation of the Ising model with spin-exchange Kinetics. In a single step of MC dynamics, a randomly selected spin $S_i$ is exchanged with a randomly chosen nearest-neighbor $S_j$ ($S_i\leftrightarrow S_j$). The change in energy $\Delta H$ that would occur if the spins were exchanged is computed. The step is then accepted or rejected with then Metropolis acceptance probability\cite{bh88, nb99}: 
\begin{eqnarray}
  P=\begin{cases}
    \exp(-\beta\Delta H) & \mathrm{for} \quad \Delta H \geq 0,  \\
    1  & \mathrm{for} \quad \Delta H \leq 0.
    \end{cases}
    \label{P}
\end{eqnarray}
Here, $\beta=(k_BT)^{-1}$ denotes the inverse temperature; $k_B$ is the Boltzmann constant. One Monte Carlo step (MCS) is completed when this algorithm is performed $N$ times (where $N$ is the total number of spins), regardless of whether the move is accepted or rejected. Noticeably, if at least one of the spin in the randomly chosen spin pair belongs to the disordered site, the proposed spin exchange will be accepted. 

The morphology of the evolving system is usually characterized by studying the two-point ($\vec{r}=\vec{r_1}-\vec{r_2}$) equal-time correlation function:
\begin{equation}
\label{cf}
C(\vec{r},t) = \frac{1}{N} \sum_{i=1}\left[\langle S_i(t) S_{i+\vec{r}}(t)\rangle - \langle S_i(t)\rangle \langle S_{i+\vec{r}}(t)\rangle \right],
\end{equation}
which measures the overlap of the spin configuration at distance $(\vec{r}$). Here, the angular brackets denote an average over different initial configurations and different noise realizations. However, most experiments study the structure factor, which is the Fourier transform of the correlation function,
\begin{equation}
\label{sf}
 S(\vec{k},t)=\sum_{\vec{r}} \exp(i\vec{k}\cdot\vec{r}) C(\vec{r},t),
\end{equation}
where $\vec{k}$ is the scattering wave-vector. Since the system under consideration is isotropic, we can improve statistics by spherically averaging the correlation function and the structure factor. The corresponding quantities are denoted as $C(r,t)$ and $S(k,t)$, respectively, where $r$ is the separation between two spatial points and $k$ is the magnitude of the wave-vector.

It is now a well-established fact that the domain coarsening during phase separation is a scaling phenomenon. The correlation function and the structure factor exhibit the dynamical scaling form \cite{ab94, pwphi9}
\begin{align}
\label{dsc}
 C(r,t) &= g[r/L(t)], \\
 S(k,t) &= L(t)^d f[kL(t)].
\end{align}
Here, $g(x)$ and $f(p)$ are the scaling functions. The characteristic length scale $L(t)$ (in the units of lattice spacing) is defined from the correlation function as the distance over which it decays to  (say) zero or any fraction of its maximum value [$C(r=0,t)=1$]; we find that the decay of $C(r,t)$ to 0.1 gives a good measure of average domain size $L(t)$. There are few different definitions of the length scale, but all these are equivalent in the scaling regime \textit{i.e.}, they differ only by constant multiplicative factors \cite{smv2011, op87}.

\section{Numerical results}
\label{Nr}
Using our MC simulations, we present results for the structure and dynamics of phase separating symmetric binary mixture ($50\%A$ and $50\%B$) with the bond disorder. We discuss both the cases of introducing the disorder (Case1: at randomly selected sites, and Case2: at regularly selected sites). The simulations are performed on a system of $N=L_x\times L_y$ particles of type $A$ and $B$ confined to a square lattice ($d=2$, $L_x=L_y=512$) such that the number density $\rho=1.0$. We quench the system from high-temperature homogeneous phase to a temperature $T=1.0$ ($T<T_c$) and then monitor the evolution of the system at various Monte Carlo steps. In presenting these results, our purpose is two-fold: first, we analyze the effects of bond-disorder on the domain coarsening and how the number of disordered sites ($N_1$) influences the characteristic features of the domains morphology and scaling behavior. Secondly, we intend to study how the different ways of introducing the same disorder affect phase separating kinetics in the system.

\subsection{Disorder at randomly selected sites}
\label{case1}
\begin{figure}[t]
\centering
\includegraphics[width=0.8\textwidth]{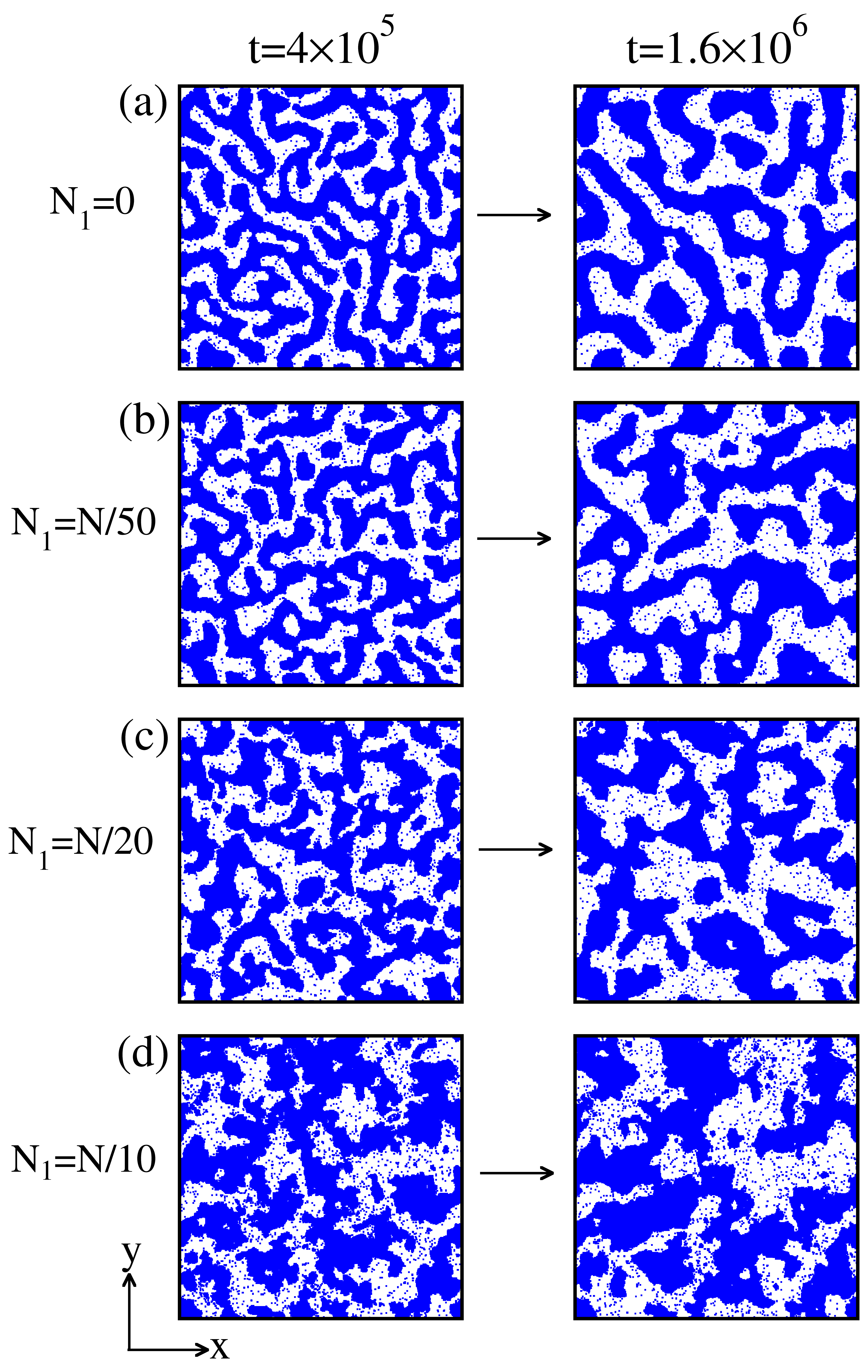}
\caption{Snapshots at $t=4\times10^5$ and $t=1.6\times10^6$ MCS, exhibiting the domain coarsening for four different percentages of disordered sites (a) 0\% ($N_1 = 0$; pure case), (b) 2\% ($N_1 = N/50$), (c) 5\% ($N_1 = N/20$), and (d) 10\% ($N_1 = N/10$). The disorder is introduced at randomly selected sites. The numerical details of the simulations are described in the text.}
\label{fig1}
\end{figure}

We present evolution morphologies of $AB$ mixture obtained from our MC simulations for Case1 in Fig. \ref{fig1} at $t=4 \times 10^5$ and $t=1.6 \times 10^6$ MCS. Fig. \ref{fig1} display the evolution pictures for four different percentages of disordered sites: (a) 0\% ($N_1 = 0$; pure case), (b) 2\% ($N_1 = N/50$), (c) 5\% ($N_1 = N/20$), and (d) 10\% ($N_1 = N/10$), respectively. Immediately after the quench, the system starts evolving via the emergence and growth of domains, namely A-rich (marked in blue) and B-rich (unmarked) regions. As expected, for a symmetric (critical) composition, a bicontinuous domain structure is seen for the pure case (Fig. \ref{fig1}a). However, with the increase of disordered sites ($N_1$), the roughening of domain walls increases \cite{hh85}; this is because of the disordered sites at which all the proposed spin exchanges are accepted and hence, domains look more fuzzier with increasing $N_1$. 

\begin{figure}[t]
\centering
\includegraphics[width=0.99\textwidth]{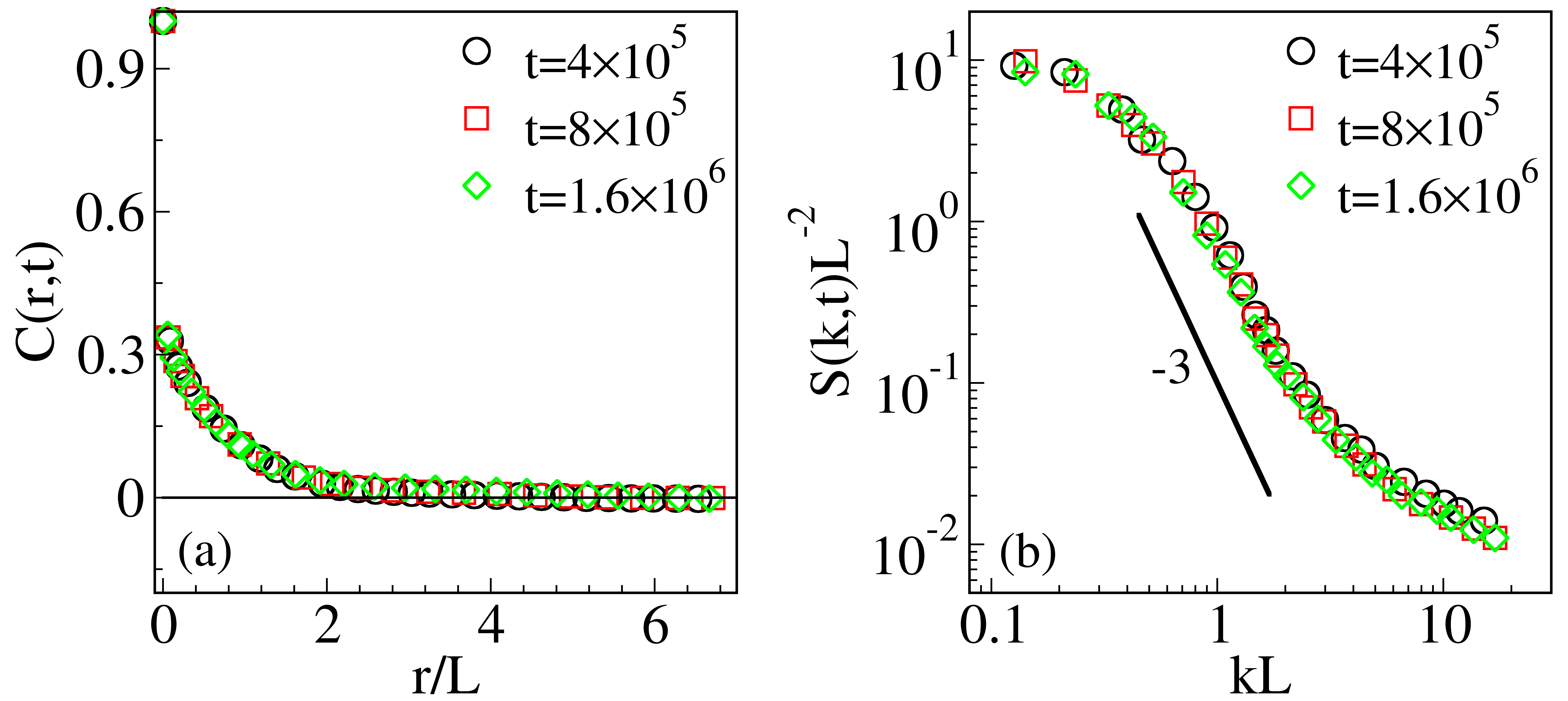}
\caption{(a) Scaling plot of $C(r,t)$ vs. $r/L$ when 5\% of randomly selected disordered sites are present in the system. The data sets at $t=4\times 10^5$, $t=8 \times10^5$, and $t=1.6 \times10^6$ MCS collapse nicely onto a single curve. (b) Plot of $S(k,t)L^{-2}$ vs. $kL$ corresponding to the same data sets as in (a). The large $k$ region (tail) of the structure factor deviates from the Porod's law, $S(k,t) \sim k^{-3}$ for $k\rightarrow \infty$. The correlation function and the structure factor data sets are obtained as an average over ten independent runs.}
\label{fig2}
\end{figure}

To study the domain morphology, we plot the scaled correlation function [$C(r,t)$ vs. $r/L(t)$] in Fig. \ref{fig2}a at three different times during the evolution. Here, we considered Case1 with 5\% of disordered sites (see Fig. \ref{fig1}c); $L(t)$ is defined as the distance over which $C(r,t)$ decays to $0.1$ of its maximum value ($C(0,t)=1$). A neat data collapse demonstrates the dynamical scaling of the domains morphology and confirms that the system for a given $N_1$ belongs to the same dynamical universality class. An excellent data collapse of the structure factor (log-log plot of $S(k,t)L^{-2}$ vs. $kL$ in Fig. \ref{fig2}b), obtained from the Fourier transform of the correlation function data sets presented in Fig. \ref{fig2}a, also demonstrate the dynamical scaling. However, for large $k$ values, $S(k,t)$ deviates from the well-known Porod's law, $S(k,t)\sim k^{-(d+1)}$, which results from scattering off sharp interfaces \cite{gp82, op88}. For other values of $N_1$, the correlation function and the structure factor exhibit the similar scaling behavior (not shown here).

\begin{figure}[t]
\centering
\includegraphics[width=0.99\textwidth]{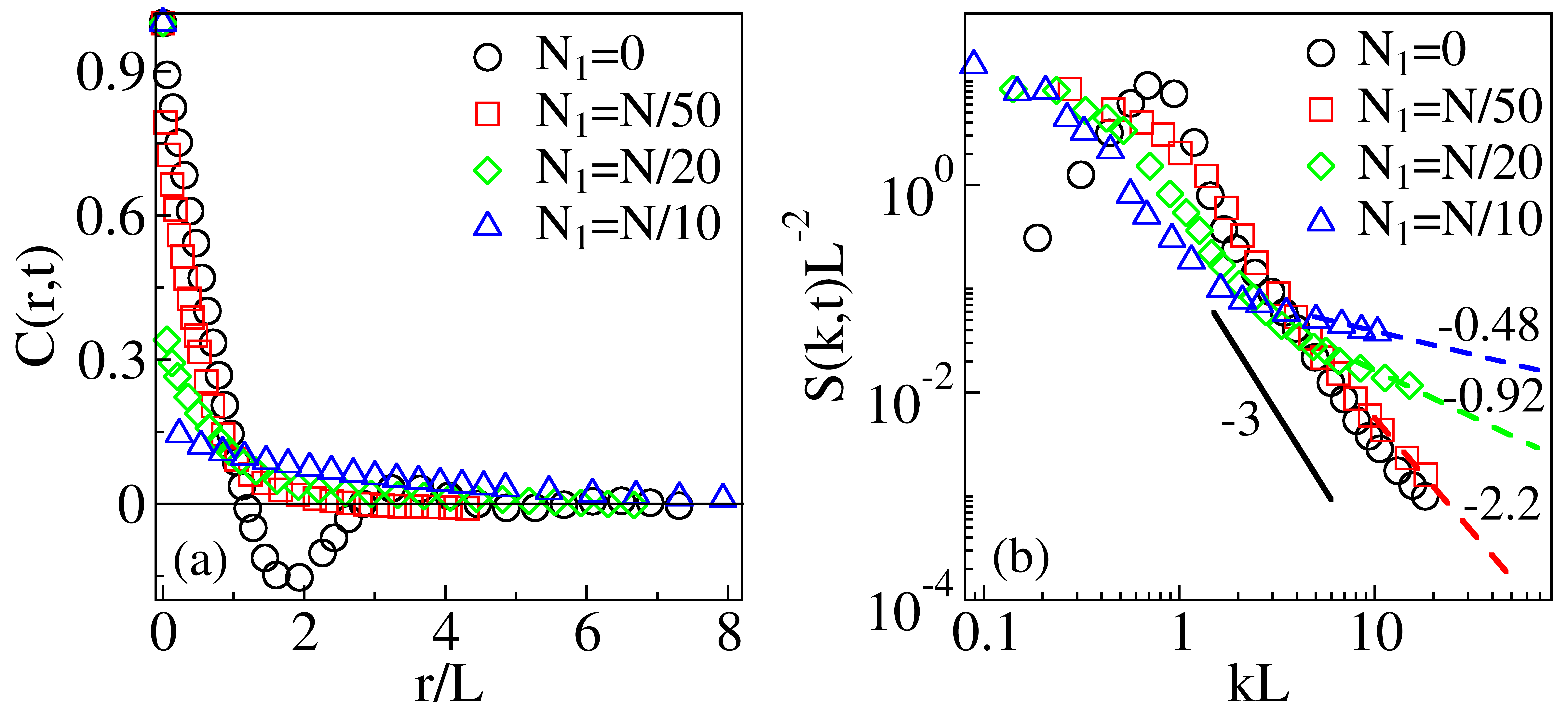}
\caption{(a) Plot of $C(r,t)$ vs. $r/L$ (for the evolution shown in Fig. \ref{fig1} at $t=1.6 \times 10^6$ MCS) at four different values of $N_1$ as denoted by the specified symbol type. With increasing $N_1$, data sets gradually deviate from the pure case (black curve) (b) Plot of $S(k,t)L^{-2}$ vs. $kL$ corresponding to the data sets in (a). For the pure case, the structure factor curve follows the Porod's law ($S(k,t) \sim k^{-3}$ for $k\rightarrow \infty$). With number of disorder sites, there is a clear deviation of the tail from the Porod's law.}
\label{fig3}
\end{figure}

We now discuss how the evolution morphology depends on the number of disordered sites, $N_1$. Fig. \ref{fig3}a shows the scaled correlation function for three different values of $N_1$ at $t=1.6 \times 10^6$ MCS when the system is already in the scaling regime (see the evolution snapshots in Fig. \ref{fig1}). The scaled correlation function for a pure binary mixture (denoted in the black symbols) is also included as a reference. Our results suggest that the data sets do not collapse onto a master function and therefore, does not belong to the same dynamical universality class. Thus, the scaling functions clearly depend upon the number of disordered sites, $N_1$. 

In Fig. \ref{fig3}b, we present the scaling plot of the structure factor, $S(k,t)L^{-2}$ vs. $kL$ on a log-log scale, corresponding to the data sets in Fig. \ref{fig3}a. For the pure system, the structure factor tail obeys the Porod's law, $S(k,t)\sim k^{-(d+1)}$ (indicated by the black symbols) as there are large regions of pure phases separated by sharp interfaces \cite{gp82, op88}. A black solid line shows the slope (-3) of the structure factor tail. The structure factor data at three different values of $N_1$ =2\%, 5\%, and 10\% are demonstrated by the red, green, and blue curves, respectively. Corresponding slopes of the structure factor tail are $-2.2$ (red dashed line), $-0.92$ (green dashed line), and $-0.48$ (blue dashed line), respectively. A deviation of the structure factor tail from the Porod's law to a lower noninteger exponent suggest a fractal architecture in the domains or interfaces as a consequence of interfacial roughening caused by quenched disorder \cite{wb88a, wb88b, sp14}. Notice that the structure factor peak shifted to a lower $k$ values with increasing $N_1$ that correspond to a large-scale structure in the system which is evident in Fig. \ref{fig1}d.  This further confirms the $N_1$ dependent scaling functions.

\begin{figure}[t]
\centering
\includegraphics[width=0.99\textwidth]{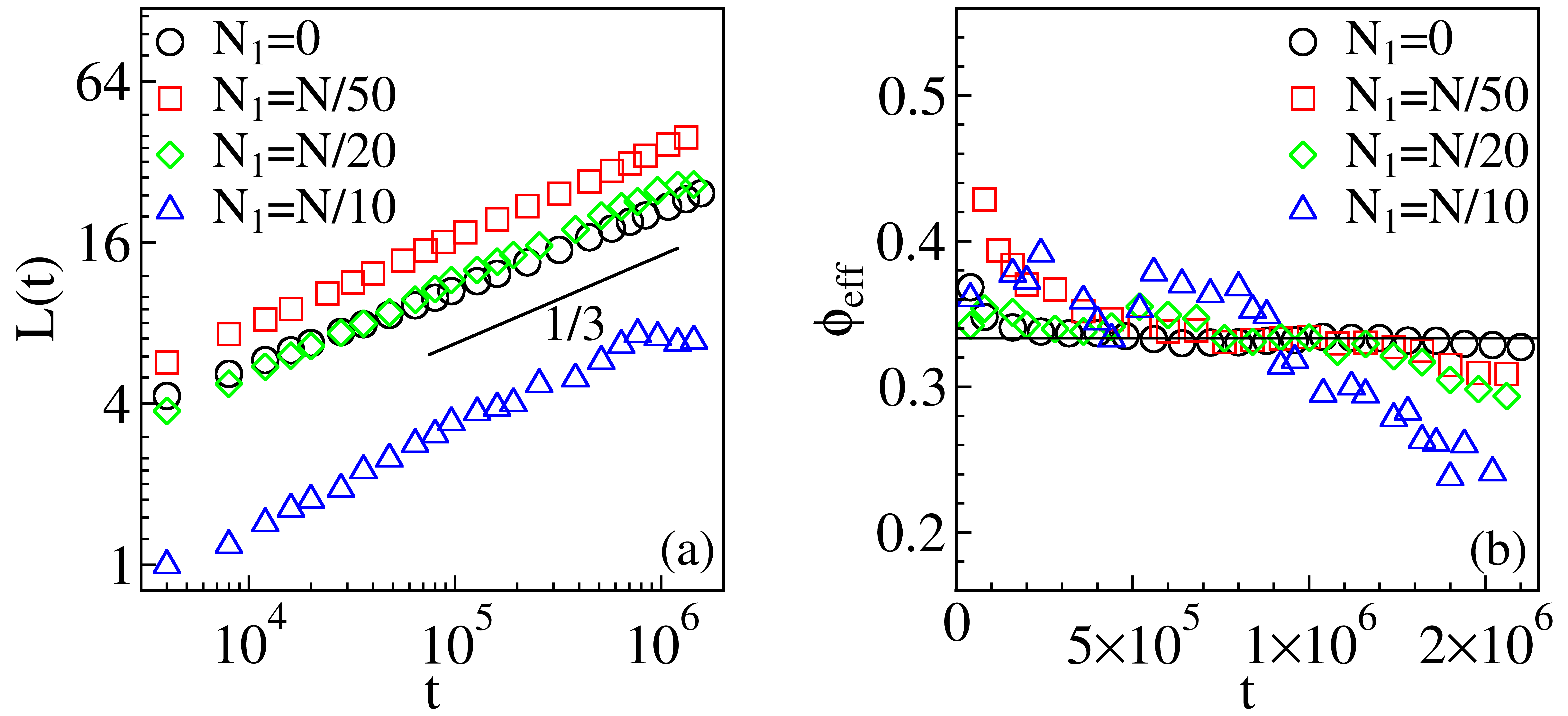}
\caption{(a) Log-log plot of the time dependence of the characteristic length scale for the evolution shown in Fig. \ref{fig1}. The symbol types represent different percentages of disorder sites. The solid black line shows the expected growth exponent $\phi=1/3$ for the pure binary mixture. (b) Variation of the effective exponent with time for the data shown in (a).}
\label{fig4}
\end{figure}

The results of the time dependence of average domain size $L(t)$ vs. $t$ are displayed in Fig. \ref{fig4} for the morphologies shown in Fig.~\ref{fig1}. For the pure case ($\epsilon=0$), coarsening morphology follows the standard Lifshitz-Slyozov (LS) growth law: $L(t)\sim t^{1/3}$ (black symbols); the black solid line represent the expected growth exponent ($\phi=1/3$) in Fig. \ref{fig4}a. For all values of $N_1 \neq 0$, our data clearly follows the LS growth law for an extended period, although, the prefactors of the power-law growth varies with $N_1$. However, on the time scale of our simulation, the domain growth law for $N_1=10\%$ crosses over to the saturation beyond $t>10^6$, which is a sign of the presence of frozen morphologies. 

Another concurrent way of extracting the growth law exponent is to define an effective growth exponent \cite{dh86, cg89},
\begin{align}
\label{phi_eff}
 \phi_{eff}(t)= \frac{\log_{\alpha}L(\alpha t)}{\log_{\alpha}L(t)}.
\end{align}
Here, we chose $\alpha=10$ \cite{cg89}. The corresponding plots of time variation of $\phi_{eff}$ are shown in Fig. \ref{fig4}b. Notice that in the pure case ($\epsilon=0$), numerical growth exponent data consistent with LS growth exponent. However, for other cases, the asymptotic growth exponents slightly deviate from the expected values with $N_1$.

Overall, we find that the system with the disorder at randomly selected sites follows the expected LS power-law growth: $L(t)\sim t^{\phi}$ with $\phi = 1/3$. For a fixed number of disorder sites ($N_1$), the system displayed the dynamical scaling at various time steps. However, the system deviates significantly from the dynamical scaling for different $N_1$ values at a fixed time step. 

\subsection{Disorder at regularly selected sites}
\label{case2}
We now examine Case2, where the disorder is introduced at the regularly selected sites by keeping the other numerical details same as in Case1. The entire system consists of $N=L_x \times L_y$ sites. The set of indexes $i = 1 \cdots L_x$ and $j = 1 \cdots L_y$, defines the respective positions of the sites in $x$, and $y$ directions. We sweep the entire lattice sites ($N$) by tracing all the indexes in $y$-direction ($1 \cdots Ly$) at each fixed $i$. In the process, every $m^{\text{th}}$ site is selected to introduce the quenched disorder. The total number of disordered sites in the system are $N_1=N/m$. We investigate the domain morphologies and the corresponding scaling properties by varying the number of disordered sites ($N_1$) and compare them with the pure case ($\epsilon=0$) as described for the Case1. 

\begin{figure}[t]
\centering
\includegraphics[width=0.8\textwidth]{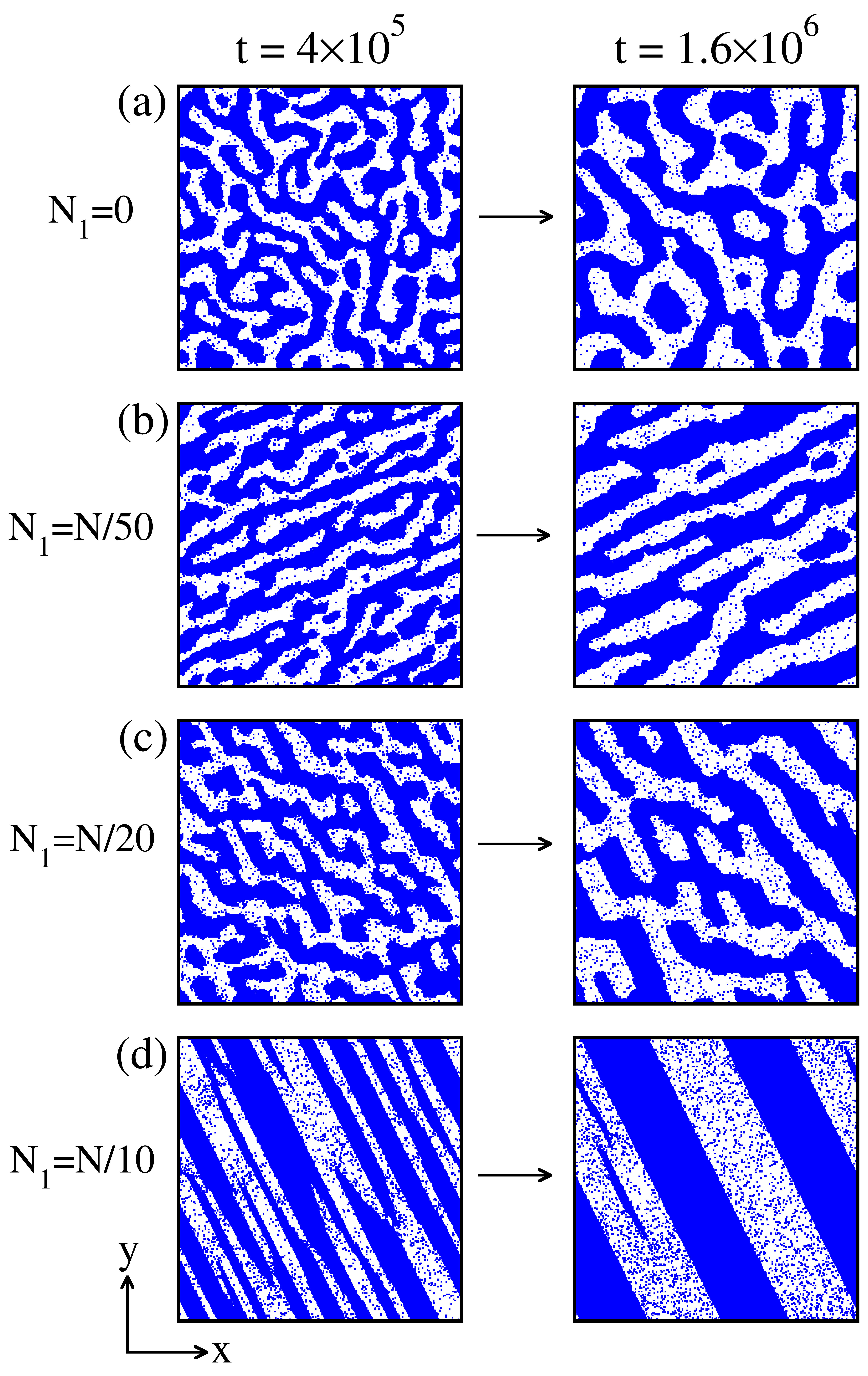}
\caption{Snapshots at $t=4\times10^5$ and $t=1.6\times10^6$ MCS, for various percentages of disorder sites in the system ($N_1$): (a) 0\%, (b) 2\%, (c) 5\%, and (d) 10\%. The disorder is introduced at regularly selected sites. Other numerical details of the simulations are described in the text.}
\label{fig5}
\end{figure}

Fig. \ref{fig5} shows the evolution morphologies at $t=4 \times 10^5$ and $t=1.6 \times 10^6$ MCS for the number of disordered sites (a) $N_1= 0$ ($0\%$), (b) $N_1= N/50$ ($2\%$), (c) $N_1= N/20$ ($5\%$), and (d) $N_1= N/10$ ($10\%$), respectively. After the temperature quench, $A$-rich (marked in blue) and $B$-rich (unmarked) domains started growing with the passage of time. In this process, where we select disordered sites in a regular manner,  stripped pattern morphology is observed (see Fig. \ref{fig5}b-d). In Fig. \ref{fig5}d, we find that with $N_1=10\%$ the evolution of stripped pattern resulting in a lamellar pattern at late times. Furthermore, we believe that even with a lower number of disorder sites ($N_1=2\%$ and $5\%$) lamellar pattern could be observed at late times $t\gg1.6\times 10^6$ MCS (see Fig. \ref{fig5}b-c), whereas such lamellar patterns occurred earlier for $N_1=10\%$. The reason for the stripped pattern could be due to the melting of domains near the regularly chosen disordered sites for which $J=0$. Hence the evolution of such systems leads to the stripe/lamellar pattern formations. Fig. \ref{fig5}b-d also reveals the dependence of stripe orientation on the number of disordered sites, $N_1$. Thus, by the combination of phase separation phenomenon of a binary mixture and the introduction of disorder at the regularly selected sites, one can guide the typical morphology of the coexisting $A$ and $B$ phases into an ordered stripped/lamellar pattern.

\begin{figure}[t]
\centering
\includegraphics[width=0.99\textwidth]{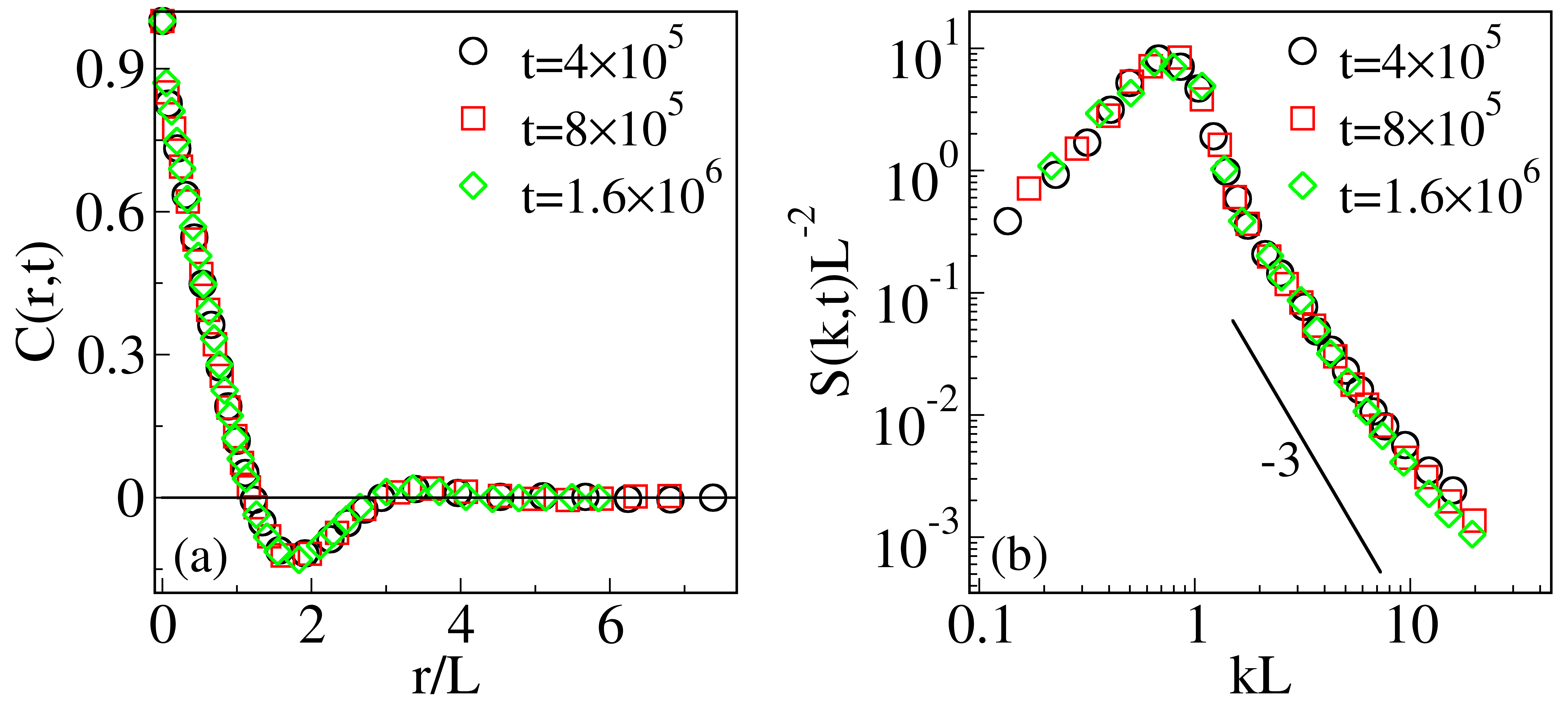}
\caption{(a) Scaling plot of $C(r,t)$ vs. $r/L$ with $N_1=5\%$ at regularly selected sites. The data sets at $t=4\times 10^5$, $t=8 \times10^5$, and $t=1.6 \times10^6$ MCS collapse nicely onto a single curve. (b) Scaling plot of $S(k,t)L^{-2}$ vs. $kL$ corresponding to the same data sets as in (a). The structure factor tail (large $k$ region) obeys the Porod's law $S(k,t) \sim k^{-3}$ for $k\rightarrow \infty$ for all the values of $N_1$ as represented by specified symbols.}
\label{fig6}
\end{figure}

Next, we present the scaling plots of the correlation function ($C(r,t)$ vs. $r/L(t)$ in Fig. \ref{fig6}a) and the structure factor ($S(k,t)L^{-2}$ vs. $kL$ in Fig. \ref{fig6}b), defined in Eq. (\ref{dsc}). Fig. \ref{fig6} corresponds to the morphologies shown in Fig. \ref{fig5}c with 5\% disordered sites. We plot the scaling functions at three time instants as indicated by the symbols. The dynamics regarding the correlation function and the structure factor at different times has shown a perfect congruence with each other witnessing the universality in their behavior as well as confirming the validity of dynamical scaling. We also observed that unlike the previous case, here the structure factor data obeys the Porod's law ($S(k,t)\sim k^{-3}$ as $k \rightarrow\infty$) which results from scattering off sharp interfaces.  

\begin{figure}[t] 
\centering
\includegraphics[width=0.99\textwidth]{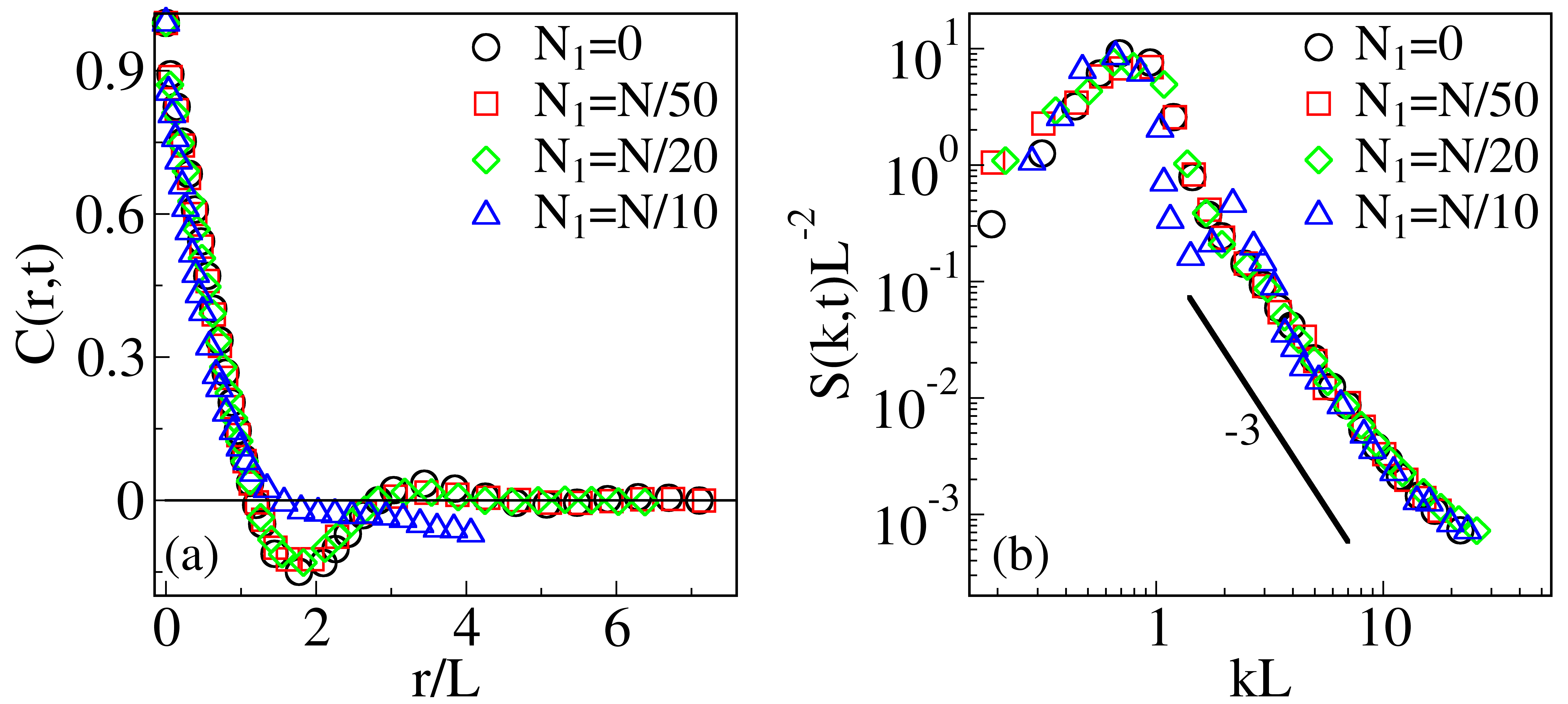}
\caption{(a) Plot of $C(r,t)$ vs. $r/L$ for the evolutions shown in Fig. \ref{fig5} at $t=1.6 \times10^6$. Except at $N_1=10\%$, we observe a good data collaps for other percentages of disorder sites (0\%, 2\% and 5\%). (b) Plot of $S(k,t)L^{-2}$ vs. $kL$ corresponding to the data sets in (a). For all the cases, tail of the structure factor obeys the Porod's law $S(k,t) \sim k^{-3}$ for $k\rightarrow \infty$. The correlation function and the structure factor data sets are obtained as an average over ten independent runs.}
\label{fig7}
\end{figure}

We now discuss whether the evolution morphology depends on the number of disordered sites present in the system. Fig. \ref{fig7} shows a comparison of the scaled correlation function and the corresponding structure factor at four different percentages of disordered sites ($N_1=0\%$, 2\%, 5\% and 10\%) for $t=1.6\times 10^6$ MCS. At lower values of $N_1$, particularly at (2\% and 5\%), excellent data collapse with the pure case ($N_1=0\%$) suggest that they belong to the same dynamical universality class i.e. the morphologies are equivalent and their statistical properties are independent of $N_1$. However, for $N_1=10\%$ the interconnected morphology of $A$ and $B$ phases transformed into an ordered lamellar pattern, hence the deviation from the dynamical scaling. Notice that the scaled correlation function for $N_1=10\%$ (shown by the blue symbols) in Fig. \ref{fig7} exhibits a crossover due to the formation of lamellar morphology. In Fig. \ref{fig7}b, the structure factor data sets also manifest the excellent data collapse on the master curve for $N_1=0\%$, 2\%, 5\%. However, notice that the structure factor for $N_1=10\%$ shows a distinct shoulder, which characterizes the lamellar structure in Fig. \ref{fig5}d. The scaled structure factor shows a Porod tail $S(k,t)\sim k^{-(d+1)}$ as $k \rightarrow\infty$ for all the values $N_1$.

\begin{figure}[t]
\centering
\includegraphics[width=0.99\textwidth]{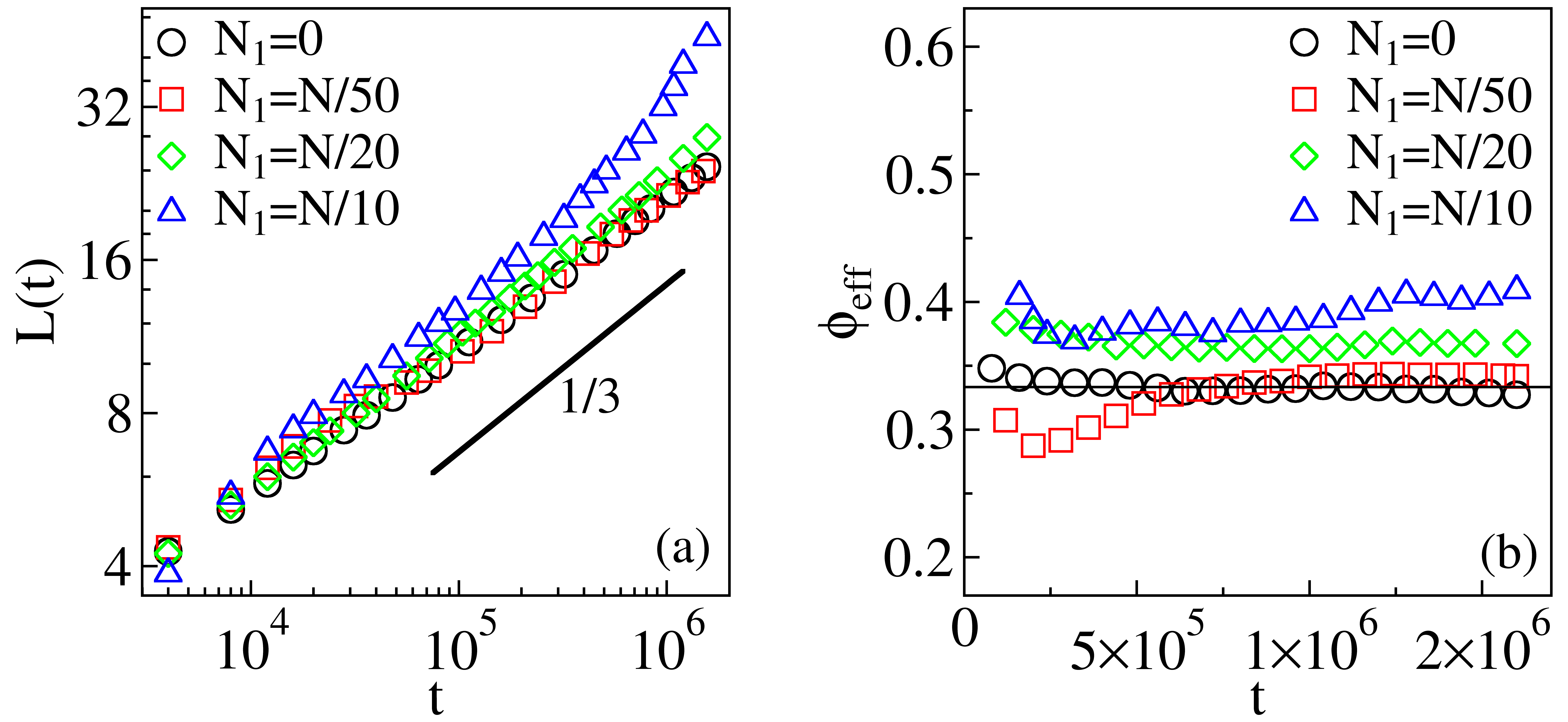}
\caption{(a) Log-log plot of the time dependence of the characteristic length
scale for the evolution shown in Fig. \ref{fig5}. The symbol types represent the number of disorder sites. The line of slope $1/3$ corresponds to the expected growth regime for pure binary mixture. (b) Variation of the effective growth exponent with time for the data shown in (a).}
\label{fig8}
\end{figure}

Finally, we turn our attention to the time dependence of domain size for the evolution shown in Fig. \ref{fig5}. We plot $L(t)$ vs. $t$ on a log-log scale in Fig. \ref{fig8}a for various $N_1$ values. The corresponding plots of $\phi_{eff}$ vs. $t$ are shown in Fig. \ref{fig8}b. We find that, after an initial transient, our data is consistent with the power-law growth for all the percentages of disorder introduced at regularly selected sites. The slight upward trend of the curves for $N_1\neq 0$ in the log-log plot suggest that the growth cannot be slower than a power-law growth. This is verified in Fig. \ref{fig8}b where we show that the variation of effective growth exponent with the number of disordered sites.

\section{Conclusions}
\label{Conclusions}
We have undertaken extensive Monte Carlo simulations to study the segregation kinetics in binary mixtures with bond-disorder. Our studies are based on kinetic Ising model with the conserved (Kawasaki) spin-exchange dynamics. We presented results for two different cases of introducing bond-disorder in the system: (i) at randomly selected sites, and (ii) at regularly selected sites, where the exchange interaction $J=1-\epsilon$ with $\epsilon =1$ and remaining sites have $\epsilon = 0$. We discussed the characteristic features of domains morphologies of phase separating ($AB$) mixtures with critical composition (50\% $A$ and 50\% $B$) for a broad range of percentages of the disorder sites $N_1=$ 0\%, 2\%, 5\%, and 10\%. 

When the disorder is incorporated at randomly selected sites (Case1), the scaling functions $C(r,t)$ and $S(k,t)$ appear to be dependent on the number of disordered sites. We observe that the domain growth law is always consistent with the Lifshitz-Slyozov (LS) growth law. However, on the time scale of our simulation, the data for a higher number of disorder sites (10\%) have crossed over to a saturation regime. We have not accessed this crossover regime for lower percentages of disorder sites; nevertheless, we cannot rule out the possibility of saturation of growth law at even later times than those investigated here. 

Next, in the Case2 where we introduced disorder at sites selected in a regular manner, evolution morphologies lead to a stripped/lamellae pattern formation. In this case, for the lower percentages (2\% and 5\%) of disorder sites, domains morphologies, which are mostly connected stripes, showing a good scaling behavior. Whereas for $10\%$ disordered sites, these system does not fall into the same universality class as the morphology is now a lamellar pattern. Hence, we observed a corresponding crossover in the scaling functions. The domain growth law, in this case, is also consistent with LS growth law on the time scale of our simulation as in the Case1. 

Overall, we believe that the results presented here will provoke a fresh interest in this significant problem, particularly, the experimental studies on the kinetics of phase separation in disordered binary mixtures.

\section{Acknowledgements}
A.S. is thankful to Prof. Sanjay Puri, SPS, JNU, India for the fruitful discussion and providing the computational facilities. A.S. is grateful to CSIR, New Delhi for the financial support. A.C. acknowledges the financial support from grant number BT/BI/03/004/2003(C) of  Government of India Ministry of Science and Technology, Department of Biotechnology, Bioinformatics division, and DST-PURSE grant given to JNU by the Department of Science and Technology, Government of India.

\newpage

\clearpage

\clearpage

\clearpage

\clearpage

\clearpage

\clearpage

\clearpage

\clearpage

\end{document}